\documentclass{article}

\usepackage{amsmath}
\usepackage{amssymb}
\usepackage{mathrsfs,dsfont}
\usepackage{mathrsfs,dsfont,graphicx}
\usepackage{bbm}
\usepackage{times}
\usepackage{bm}
\usepackage{natbib}
\allowdisplaybreaks
\usepackage[plain,noend]{algorithm2e}
\usepackage{soul,color}
\usepackage{a4wide}

\def\bSig\mathbf{\Sigma}

\newcommand{\Qbb}{\mathbb{Q}}

\newcommand{\xb}{\mathbf{x}}

\newcommand{\tb}{\mathbf{t}}
\newcommand{\nb}{\mathbf{n}}
\newcommand{\ub}{\mathbf{u}}
\newcommand{\vb}{\mathbf{v}}

\newcommand{\Xb}{\mathbf{X}}
\newcommand{\Yb}{\mathbf{Y}}
\newcommand{\xib}{\boldsymbol{\xi}}
\newcommand{\thetab}{\boldsymbol{\theta}}
\newcommand{\ud}{\mathrm{d}}

\newcommand{\pp}{\textsf{P}}
\newcommand{\ee}{\textsf{E}}

\newcommand{\samplespace}{\mathbb{X}}
\newcommand{\samplesigmafield}{\mathscr{X}}
\def\rd{\mathbb{R}^d}

\def\naturals{\mathbb{N}}
\def\reals{\mathbb{R}}
\newcommand{\ft}{f(\cdot;\thetab)}
\newcommand{\TETA}{\thetab \in \Theta}

\makeatletter
\renewcommand{\algocf@captiontext}[2]{#1\algocf@typo. \AlCapFnt{}#2} 
\def\@algocf@capt@plain{top}
\renewcommand{\algocf@makecaption}[2]{%
  \addtolength{\hsize}{\algomargin}%
  \sbox\@tempboxa{\algocf@captiontext{#1}{#2}}%
  \ifdim\wd\@tempboxa >\hsize
    \hskip .5\algomargin%
    \parbox[t]{\hsize}{\algocf@captiontext{#1}{#2}}
  \else%
    \global\@minipagefalse%
    \hbox to\hsize{\box\@tempboxa}
  \fi%
  \addtolength{\hsize}{-\algomargin}%
}
\makeatother

\newtheorem{theorem}{Theorem}

\newtheorem{lemma}[theorem]{Lemma}

\begin{document}




\title{A Wilks' theorem for grouped data}

\date{}

\author{Emanuele Dolera \\
\emph{University of Pavia} \\
email: emanuele.dolera@unipv.it \\
\\
Stefano Favaro \\
\emph{University of Torino and Collegio Carlo Alberto} \\
email: stefano.favaro@unito.it \\
\\
Andrea Bulgarelli \\
\emph{INAF-IASF Bologna} \\
email: bulgarelli@iasfbo.inaf.it \\
\\
Alessio Aboudan \\
\emph{CISAS, University of Padova} \\
email: alessio.aboudan@unipd.it
}

%
%
%
\maketitle

\begin{abstract}
Consider $n$ independent measurements, with the additional information of the times at which measurements are performed. This paper deals with testing statistical hypotheses when $n$ is large and only a small amount of observations concentrated in short time intervals are relevant to the study. We define a testing procedure in terms of multiple likelihood ratio (LR) statistics obtained by splitting the observations into groups, and in accordance with the following principles: P1) each LR statistic is formed by gathering the data included in $G$ consecutive vectors of observations, where $G$ is a suitable time window defined a priori with respect to an arbitrary choice of the ``origin of time''; P2) the null statistical hypothesis is rejected only if at least $k$ LR statistics are sufficiently small, for a suitable choice of $k$. We show that the application of the classical Wilks' theorem may be affected by the arbitrary choice of the ``origin of time", in connection with P1). We then introduce a Wilks' theorem for grouped data which leads to a testing procedure that overcomes the problem of the arbitrary choice of the ``origin of time'', while fulfilling P1) and P2). Such a procedure is more powerful than the corresponding procedure based on Wilks' theorem.
\end{abstract}



\section{Introduction}

Consider $n$ independent measurements of the same physical phenomenon, with the additional information of the times at which measurements are performed. This paper deals with the problem of testing statistical hypotheses when $n$ is large and only a small amount of observations concentrated in short time intervals (critical phenomena) are relevant to the study under investigation. The sample consists of $\{(X_1, t_1), \dots, (X_n, t_n)\}$, where $(X_1, \dots, X_n)$ are independent and identically distributed (iid) random variables, and $(t_1, \dots, t_n)\in(0, +\infty)^n$. We make use of $(t_1, \dots, t_n)$ to split $(X_1, \dots, X_n)$ into $P$ vectors, as follows: fix a basic unit of time in such a way that the whole dataset corresponds to the observation of $P$ units of time, and define the random vector $\mathbf{X}^{(p)}:= (X^{(p)}_1, \dots, X^{(p)}_{n_p})$, for $p = 1, \dots, P$, whose $n_{p}$ components are the $X_i$'s such that $t_i \in (p-1, p]$. For instance, for a phenomenon with measurements at every minute for multiple years, $P$ may be the number of hours in a year. To complete the picture, let $\samplespace$ be the set of all possible realizations of any trial, endowed with the  $\sigma$-algebra $\samplesigmafield$, and consider a regular parametric model $\{\ft;\TETA\}$, where $\Theta$ is an open subset of $\rd$ and, for every $\thetab \in \Theta$, $x \mapsto f(x;\thetab)$ is a probability density function with respect to a $\sigma$-finite reference measure $\nu$ on $(\samplespace, \samplesigmafield)$. The notion for regular parametric models will be made precise in Section \ref{sec1}. The common probability density function of the $X^{(p)}_j$'s is denoted by $f(\cdot;\thetab_0)$, where $\thetab_0 \in \Theta$ is the true, but unknown, value of the parameter $\thetab$. The objective of our study is to test the null hypothesis $H_0 : \thetab_0 \in \Theta_0$ against the alternative hypothesis $H_1 : \thetab_0 \not\in \Theta_0$, where $\Theta_0$ denotes a proper subset of $\Theta$.

We define a testing procedure in terms of multiple likelihood ratio (LR) statistics, and in accordance with the following principles: P1) each LR statistic is formed by gathering observations included in $G$ subsequent vectors $\mathbf{X}^{(p)}$'s, i.e. observations $X_{i}$'s whose $t_{i}$'s belong to $G$ subsequent units of time, where $G$ is a suitable time window defined a priori with respect to an arbitrary choice of the ``origin of time"; P2) $H_0$ is rejected only if at least $k$ LR statistics are sufficiently small, for a suitable choice of $k$. The time window $G$ in P1) allows for tuning the LR statistics with respect to what, a priori, is considered to be the typical duration (length of time intervals) of critical phenomena that are supposed to be induced by $H_1$. Then P2) is justified whenever it is desirable to have repeated manifestations of critical phenomena to accept $H_1$. Our testing procedure with multiple LR statistics based on P1) and P2) is motivated by the above premise that, among a large number of observations, only critical phenomena are relevant to the study. Indeed, if relevant observations  are concentrated in short time intervals of duration less than $G$ units of time, then the analysis based  on a single LR statistic would be meaningless, since the overwhelming majority of observations would always lead to accept $H_0$. On the contrary, the application of P1), in conjunction with a reasonable choice of the time window $G$, ensures that observations may be relevant with respect to a subgroup of observations detected during a period of $G$ units of time.

Wilks' theorem on large sample asymptotics for LR statistics (\cite{Wil(38)} and \cite{Wal(43)}) may be applied to devise a testing procedure fulfilling principles P1) and P2). Let $P = N \cdot G$, with $N$ being the number of LR statistics, and let $\xb^{(1)} := (x^{(1)}_1, \dots, x^{(1)}_{n_1}), \dots, \xb^{(P)} := (x^{(P)}_1, \dots, x^{(P)}_{n_P})$ be the collected data. Then we define the vector of LR statistics $\mathbf{\Lambda}^{(st)}:=(\Lambda_1^{(st)}, \dots, \Lambda_N^{(st)})$, where $\Lambda_i^{(st)}$ is obtained by gathering data belonging to vectors from $(i-1)G + 1$ to $iG$, i.e.
\begin{equation} \label{eq:LambdaST}
\Lambda_i^{(st)} := \Lambda_i^{(st)}(\xb^{((i-1)G + 1)}; \dots; \xb^{(iG)}) := 
\frac{\sup_{\thetab \in \Theta_0} \prod_{p = (i-1)G + 1}^{iG} \prod_{j = 1}^{n_p} f(x_j^{(p)};\thetab)}{\sup_{\thetab \in \Theta} \prod_{p = (i-1)G + 1}^{iG}
\prod_{j = 1}^{n_p} f(x_j^{(p)};\thetab)}
\end{equation}
for $i = 1, \dots, N$. Note that, in this framework, the components $\Lambda_i^{(st)}(\mathbf{X}^{((i-1)G + 1)}; \dots; \mathbf{X}^{(iG)})$ of $\mathbf{\Lambda}^{(st)}$ turn out to be stochastically independent, since the $N$ groups of $G$ vectors just considered are disjoint. Then, reject $H_0$ if at least $k$ of the $\Lambda_i^{(st)}$'s are less than some reference value $\alpha$. Due to independence, the probability of type I error can be evaluated by means of the binomial formula as $\pi(k, \alpha) := \sum_{k\leq h \leq N}\binom{N}{h} p_{\alpha}^h (1 - p_{\alpha})^{N-h}$, where the probability $p_{\alpha}$ that a single $\Lambda_i^{(st)}$ is less than $\alpha$ can be approximated, with sufficiently good precision, by resorting to Wilks' theorem. In fact, from this theorem one has: if $\{\ft;\TETA\}$ is a regular parametric model and $\Theta_0$ is an $s$-dimensional ($s \in \{1, 2, \dots, d-1\}$) sub-manifold of  $\Theta$, then, under $H_0$, the probability distribution function of $-2 \log \Lambda_i^{(st)}(\mathbf{X}^{((i-1)G + 1)}; \dots; \mathbf{X}^{(iG)})$ converges weakly, for every $i = 1, \dots, N$, to a standard $\chi^2$ distribution with $d-s$ degrees of freedom, as $n_1, \dots, n_P$ go to infinity.

While the above testing procedure is simple and supported by Wilks' theorem, the number of LR statistics less than $\alpha$ may be affected by the arbitrary choice of the ``origin of time'', in connection with P1). Indeed since $H_1$ is supported by critical phenomena of duration less than $G$ units of time, each of these phenomena is completely seized in a LR statistic only if both its initial time and the final time belong to the interval $((i-1)G, iG]$. On the contrary, if the initial time of a critical phenomena belongs to $((i-1)G, iG]$ and the final time belongs to $(iG, (i+1)G]$, such a phenomena is not seized, or it is partially seized, with both the LR statistic $\Lambda_i^{(st)}$ and the LR statistic $\Lambda_{i+1}^{(st)}$ being possibly greater than $\alpha$. The application of Wilk's theorem thus implies a specific choice for the ``origin of time'', unless one neglects observations belonging to units of time in between time windows. Clearly, this may affect remarkably the decision process. In this paper we propose an alternative testing procedure which overcomes the problem of the arbitrary choice of the ``origin of time'', while fulfilling principles P1) and P2). According to our procedure, for any choice of the time window $G$ it is no longer possible to neglect a critical phenomena (of duration less than $G$ units of time) starting at the time interval $((i-1)G, iG]$ and ending at the time interval $(iG, (i+1)G]$. Indeed there will always exist another time interval, in a new finer subdivision, which contains both the initial and the final time instants of the critical phenomena. The proposed approach relies on a novel Wilks' theorem for grouped data, which leads to a rejection event that includes the corresponding rejection event based on Wilks' theorem. That is, our testing procedure is more powerful than the above Wilks' testing procedure.


\section{Methodology}\label{sec1}
Consider $M$ groups of $G$ consecutive vectors, the $i$-th vector consisting of those vectors that are numbered from $i$ to $i + G - 1$, where $M = P - G + 1 = (N-1)G + 1$. Once the data are collected in the form $\xb^{(1)} := (x^{(1)}_1, \dots, x^{(1)}_{n_1}), \dots, \xb^{(P)} := (x^{(P)}_1, \dots, x^{(P)}_{n_P})$, we associate a LR statistic with each group, obtaining the vector of LR statistics $\boldsymbol{\Lambda}^{(new)} := (\Lambda_1^{(new)}, \dots, \Lambda_M^{(new)})$ defined by 
\begin{equation} \label{eq:Lambdanew}
\Lambda_i^{(new)} := \Lambda_i^{(new)}(\xb^{(i)}; \dots; \xb^{(i + G - 1)}) := \frac{\sup_{\thetab \in \Theta_0} \prod_{p = i}^{i + G - 1} \prod_{j = 1}^{n_p} f(x_j^{(p)}\ |\ \thetab)}{\sup_{\thetab \in \Theta} \prod_{p = i}^{i + G - 1} \prod_{j = 1}^{n_p} f(x_j^{(p)}\ |\ \thetab)}
\end{equation}
for $i = 1, \dots, M$. Differently from $\mathbf{\Lambda}^{(st)}$, the components $\Lambda_i^{(new)}(\mathbf{X}^{(i)}; \dots; \mathbf{X}^{(i+G-1)})$'s  of $\boldsymbol{\Lambda}^{(new)}$ are no more independent. Therefore, our testing procedure will deal with the joint probability distribution of $\boldsymbol{\Lambda}^{(new)}(\mathbf{X})$, and in particular with its asymptotic behaviour for large values of the sample sizes $n_1, \dots, n_P$. Our result will not provide weak convergence of $\boldsymbol{\Lambda}^{(new)}$ towards a specific limiting distribution, but only a merging phenomenon, in the following sense: after fixing a distance to compare probability distributions on $(\reals^M, \mathscr{B}(\reals^M))$, we will provide an approximating sequence such that the distance between the probability distribution of $\boldsymbol{\Lambda}^{(new)}(\mathbf{X})$ and the relative element of the approximating sequence goes to zero as $n_1, \dots, n_P$ go to infinity. The approximating sequence depends on the data only through the sample sizes $n_1, \dots, n_P$, and it does not depend on the model $\{\ft;\TETA\}$ and of the choice of $\thetab_0$. With such a theoretical result at disposal, we can describe a testing procedure which overcomes the problem of the arbitrary choice of the ``origin of time'' while fulfilling principles P1) and P2). Such a procedure consists of rejecting $H_0$ whenever there are at least $k$ of the $\Lambda_i^{(new)}$'s, say $\Lambda_{i_1}^{(new)}, \dots, \Lambda_{i_k}^{(new)}$, with $i_{j+1} - i_j \geq G$, which are less than $\alpha$. Formally, the rejection rule corresponds to considering the event 
\begin{displaymath}
\cup_{\substack{1 \leq i_1 < \dots < i_k \leq M \\ i_{j+1} - i_j \geq G}} \Big{(} \{\Lambda_{i_1}^{(new)} < \alpha\} \cap \dots \cap \{\Lambda_{i_k}^{(new)} < \alpha\} \Big{)}
\end{displaymath}
whose probability can be evaluated after knowing the joint probability distribution of $\boldsymbol{\Lambda}^{(new)}(\mathbf{X})$. Theorem \ref{thm:wilks} below provides with an explicit approximation of such a joint probability distribution for LR statistics.

Before stating Theorem \ref{thm:wilks}, it is worth recalling that the parametric model $\{\ft;\TETA\}$ is called \emph{regular} when the following conditions are met:
\begin{enumerate}
\item[C1)] $\forall\ x \in \samplespace$, $\thetab \mapsto f(x;\thetab)$ belongs to $C^2(\Theta)$;
\item[C2)] the set $\samplespace_+ := \{x \in \samplespace\ |\ f(x;\thetab) > 0\}$ does not depend on $\thetab$ and $\nu(\samplespace_+^c) = 0$;
\item[C3)] for any measurable function $T : \samplespace \rightarrow \reals$ satisfying $\int_{\samplespace} T(x) f(x;\thetab) \nu(\ud x) < +\infty$ for all $\thetab \in \Theta$, derivatives of first and second order (with respect to $\thetab$) may be passed under the integral sign in $\int_{\samplespace} T(x) f(x;\thetab) \nu(\ud x)$;
\item[C4)] for any $\thetab_0 \in \Theta$, there exist a measurable function $K_0 : \samplespace \rightarrow [0, +\infty]$ and $\delta_0 > 0$ such that
\begin{eqnarray}
\int_{\samplespace} K_0(x) f(x;\thetab_0) \nu(\ud x) &<& +\infty\ , \nonumber \\
\sup_{|\thetab - \thetab_0| \leq \delta_0} \Big{|} \frac{\partial^2}{\partial \theta_i \partial \theta_j} \log f(x;\thetab)\Big{|} &\leq& K_0(x) \ \ \ \ \ \forall\ x \in \samplespace, i, j = 1, \dots, d; \nonumber
\end{eqnarray}
\item[C5)] the Fisher information matrix $\mathrm{I}(\thetab) := (\mathrm{I}_{i,j}(\thetab))_{i,j = 1, \dots, d}$, given by
\begin{equation} \label{eq:fisher}
\mathrm{I}_{i,j}(\thetab) := -\int_{\samplespace} \left(\frac{\partial^2}{\partial \theta_i \partial \theta_j} \log f(x;\thetab)\right) f(x;\thetab) \nu(\ud x)\ ,
\end{equation}
is well-defined and positive definite at every value of $\thetab$;
\item[C6)] the model is identified, i.e. $\nu \left( \left\{ x \in \samplespace\ |\ f(x ;\thetab_1) \neq f(x;\thetab_2) \right\} \right) = 0$ entails $\thetab_1 = \thetab_2$.
\end{enumerate}
In addition, in order to avoid technical---but not conceptual---complications in the proofs, 
we require a maximum likelihood estimator (MLE) actually exists as a point of $\Theta$, meaning that such a MLE must coincide with a root of the likelihood equation. More formally, we assume that
\begin{enumerate}
\item[C7)] $\forall\ n \geq n_0$, there exists a measurable function $\tb_n : \samplespace^n \rightarrow \Theta$ such that
\begin{equation} \label{eq:exMLE}
\sup_{\thetab \in \Theta} \Big[\prod_{j=1}^n f(x_j;\thetab)\Big] = \prod_{j=1}^n f(x_j ;\tb_n(x_1, \dots, x_n)) \ \ \ \ \ \ \ \ \ \ \forall\ (x_1, \dots, x_n) \in  \samplespace^n\ .
\end{equation}
\end{enumerate}

To formalize the concept of approximating sequence, we must introduce a suitable distance to compare probability distributions on $(\reals^l, \mathscr{B}(\reals^l))$. See, e.g., \cite{Gib(02)} or Chapter 2 of \cite{Sen(98)} for a comprehensive treatment of distances for probability distributions. Among the various possible distances, we select the L\'evy-Prokhorov distance $D_{l}$, which is particularly meaningful with respect to our problem. Specifically, given a pair $(\mu_1, \mu_2)$ of probability measures on $(\reals^l, \mathscr{B}(\reals^l))$,
\begin{displaymath}
D_l(\mu_1; \mu_2) := \inf\{ \varepsilon > 0\ |\ \mu_1(B) \leq \mu_2(B^{\varepsilon}) + \varepsilon, \mu_2(B) \leq \mu_1(B^{\varepsilon}) + \varepsilon, \ \forall\ B \in \mathscr{B}(\reals^l) \},
\end{displaymath}
where $B^{\varepsilon} := \{x \in \reals^l\ |\ \ud(x, C) \leq \varepsilon \}$. The distance $D_l$ is often used in the context of multidimensional extensions of the Berry-Esseen estimate, being related to the concept of weak convergence of probability measures (see, e.g., Section 11.3 of \cite{Du(01)}). 

Now we can state our first result, which deals with the asymptotic normality of the vector $(\hat{\thetab}_{n_1, \dots, n_G}, \dots, \hat{\thetab}_{n_M, \dots, n_P})$ of MLE's, whose components are defined by $\hat{\thetab}_{n_i, \dots, n_{i + G - 1}} := \tb_{n_i + \dots + n_{i+G-1}}(\Xb^{(i)}; \dots; \Xb^{(i+G-1)})$, for $i = 1, \dots, M$, with the same $\tb_n$ as in \eqref{eq:exMLE}.
\begin{theorem}\label{thm:cramer}
Let $ \thetab_0$ be the true, but unknown, value of $\thetab\in\Theta$, and let the conditions of regularity C1)-C7) for the parametric model $\{\ft;\TETA\}$ be satisfied. Then, the probability distribution $\mu_{n_1, \dots, n_P}^{(dM)}$ of 
$$
\left(\sqrt{\sum_{k = 1}^G n_k} \cdot (\hat{\thetab}_{n_1, \dots, n_G} - \thetab_0), \dots, \sqrt{\sum_{k = M}^P n_k} \cdot (\hat{\thetab}_{n_M, \dots, n_P} - \thetab_0)\right)\ ,
$$
meets
\begin{equation} \label{eq:cramer}
D_{dM}\left(\mu_{n_1, \dots, n_P}^{(dM)}; \gamma^{(dM)}(\mathrm{R}_M, \mathrm{I}(\thetab_0)^{-1}) \right) \rightarrow 0
\end{equation}
as $\ n_1, \dots, n_P \rightarrow +\infty$, where:
\begin{itemize}
\item[i)] $\mathrm{R}_M := \mathrm{R}_M(n_1, \dots, n_P)$ is the $M \times M$ matrix whose elements $\rho_{i,j} := \rho_{i,j}(n_1, \dots, n_P)$ are given by
\begin{equation} \label{eq:rhoij}
\rho_{i,j}(n_1, \dots, n_P) := \left\{ \begin{array}{ll}
0 & \text{if} \ i,j \in \{1, \dots, M\}, |i-j| \geq G \\
\frac{\sum_{p = a(i,j)}^{b(i,j)} n_p}{\sqrt{\sum_{q = i}^{i+G-1}\sum_{l = j}^{j+G-1} n_q n_l}} & \text{if} \ i,j \in \{1, \dots, M\}, |i-j| < G
\end{array} \right.
\end{equation}
with $a(i,j) := \max\{i, j\}$ and $b(i,j) := \min\{i, j\} + G - 1$; 
\item[ii)] $\mathrm{I}(\thetab_0)$ is defined by means of \eqref{eq:fisher}; 
\item[iii)] $\gamma^{(dM)}(\mathrm{R}_M, \mathrm{I}(\thetab_0)^{-1})$ is the $dM$-dimensional Gaussian probability distribution with zero means and covariance matrix
\begin{equation} \label{eq:cov0}
\left(\begin{array}{c|c|c|c}
\rho_{1,1} \mathrm{I}(\thetab_0)^{-1} & \rho_{1,2} \mathrm{I}(\thetab_0)^{-1} & \ldots & \rho_{1,M} \mathrm{I}(\thetab_0)^{-1} \\
\hline
\rho_{2,1} \mathrm{I}(\thetab_0)^{-1} & \rho_{2,2} \mathrm{I}(\thetab_0)^{-1} & \ldots & \rho_{2,M} \mathrm{I}(\thetab_0)^{-1} \\
\hline
\vdots & \vdots & \ddots & \vdots \\
\hline
\rho_{M,1} \mathrm{I}(\thetab_0)^{-1} & \rho_{M,2} \mathrm{I}(\thetab_0)^{-1} & \ldots & \rho_{M,M} \mathrm{I}(\thetab_0)^{-1}
\end{array} \right) \ .
\end{equation}
\end{itemize}
\end{theorem}

\smallskip

It is worth noticing that $\rho_{i,i} = 1$ for $i = 1, \dots, M$, and that the matrix $\mathrm{R}_M$ is positive-definite, as it coincides with the covariance matrix of the Gaussian random vector $(W_1, \dots, W_M)$
where $W_i := \sum_{j=i}^{i+G-1} E_j$ and $(E_1, \dots, E_P)$ is a vector of independent real random variables with $E_j \sim \mathcal{N}(0, n_j)$.

As a consequence of Theorem \ref{thm:cramer}, we can state the main result of the paper.

\begin{theorem}\label{thm:wilks}
Let $\Theta_0$ be an $s$-dimensional sub-manifold of $\Theta$, with $s \in \{1, \dots, d-1\}$, and let the conditions of regularity C1)-C7) for the parametric model $\{\ft;\TETA\}$ be satisfied. If 
$\Xi_i := \Xi_i(\mathbf{X}^{(i)}; \dots; \mathbf{X}^{(i+G-1)}) := -2\log(\Lambda_i^{(new)}(\mathbf{X}^{(i)}; \dots; \mathbf{X}^{(i+G-1)}))$, for $i=1,\ldots,M$, then, under $H_0$, the probability distribution $\eta_{n_1, \dots, n_P}^{(M)}$ of $(\Xi_1, \dots,  \Xi_M)$ meets
\begin{equation} \label{eq:wilks}
D_M\left(\eta_{n_1, \dots, n_P}^{(M)}; \chi^2_{M,r}(\mathrm{R}_M) \right) \rightarrow 0
\end{equation}
as $\ n_1, \dots, n_P \rightarrow +\infty$, where:
\begin{itemize}
\item[i)] $r := d-s$;
\item[ii)] $\chi^2_{M,r}(\mathrm{R}_M)$ stands for the probability distribution of the $M$-dimensional random vector
$$
\left(\sum_{h = 1}^r Z_{h; 1}^2, \sum_{h = 1}^r Z_{h; 2}^2, \dots, \sum_{h = 1}^r Z_{h; M}^2 \right)\ ;
$$
\item[iii)] the $rM$-dimensional random vector $(Z_{1;1}, \dots, Z_{r;1}, Z_{1;2}, \dots, Z_{r;2}, \dots, Z_{1;M}, \dots, Z_{r;M})$ is jointly Gaussian with zero means and covariance matrix given by
$$
\left\{
\begin{array}{ll}
\textsf{Var}(Z_{h; i}) = 1 & \text{if}\ h = 1, \dots, r\ \ \text{and}\ i = 1, \dots, M \\
\textsf{Cov}(Z_{h; i}, Z_{l; j}) = 0 & \text{if}\ h \neq l\ \ \text{and}\ i, j = 1, \dots, M \\
\textsf{Cov}(Z_{h; i}, Z_{h; j}) = 0 & \text{if} \ |i - j| \geq G\ \ \text{and}\ h = 1, \dots, r \\
\textsf{Cov}(Z_{h; i}, Z_{h; j}) = \rho_{i,j} & \text{if}\ |i - j| < G\ \ \text{and}\ h = 1, \dots, r\ . 
\end{array} \right.
$$
\end{itemize}
\end{theorem}

From a theoretical perspective, there is a clear improvement in using the new testing procedure based on Theorem \eqref{thm:wilks} rather then the standard testing procedure based on Wilk's theorem. This is because of the fact that the new rejection event includes its standard counterpart, entailing that the new testing proceure turns out to be more powerful than the standard testing procedure. Moreover, the problem of the arbitrary choice of the ``origin of time" is now definitely solved. Indeed it is not possible anymore to neglect a critical phenomena (of duration less than $G$ units of time) starting at the time interval $((i-1)G, iG]$ and ending at the time interval $(iG, (i+1)G]$, for any choice of $G$.


\section{Discussion}

We considered testing hypotheses under this setting: a large number of independent measurements of which only a small amount, concentrated in short periods, are relevant to the study under investigation. Our motivating example comes from recent works on detection of $\gamma$-ray astrophysical sources under the AGILE project (http://agile.asdc.asi.it). See, e.g., \cite{Bul(12)} and \cite{Bul(14)}. The $X_{i}$'s are associated to measurements of photons, with the information being the position of the photon in the sky and its energy. The basic unit of time is the hour, and the iid assumption is motivated by the fact that the region of the sky under investigation is invariant for the duration of the AGILE project (5 years). The dataset consists of a huge number of observations, but only a small amount of them, concentrated in periods of less than 24 hours, are relevant. Indeed the number of photons ascribable to distinguish astrophysical sources (e.g., supernova remnants, black hole binaries and pulsar wind nebulae) is much smaller than the total number of observed photons. \cite{Bul(12)} relied on the statistic \eqref{eq:LambdaST}, with $G=24$, for testing certain hypotheses related to the detection of $\gamma$-ray astrophysical sources. In this paper we discussed how \eqref{eq:LambdaST}, with an arbitrary choice of the ``origin of time", may lead to a meaningless analysis. We then introduced an alternative, and more powerful, test that allows for an arbitrary choice of the ``origin of time". Such a procedure relies on the novel Wilks' theorem for grouped data, which may be of independent interest. Since a precise formulation of the problem in \cite{Bul(12)} would require to introduce certain (technical) protocols of the AGILE project, we defer the application of our approach to a companion paper for a journal in astrophysics.


\section{Proofs}

The proofs of the main theorems are based on the following three lemmas. 
\begin{lemma} \label{lm:tight}
Let $\{\beta_n\}_{n \geq 1}$ and $\{\beta^{'}_n\}_{n \geq 1}$ be two sequences of p.m.'s on $(\reals^l, \mathscr{B}(\reals^l))$. If $\{\beta^{'}_n\}_{n \geq 1}$ is tight and $D_l(\beta_n; \beta^{'}_n) \rightarrow 0$ as $n \rightarrow +\infty$, then
$\{\beta_n\}_{n \geq 1}$ is also tight. 
\end{lemma}

\noindent \textbf{Proof of Lemma \ref{lm:tight}}.
For any $\varepsilon > 0$, denote by $\rho(\varepsilon) > 0$ a positive number such that $\sup_{n \in \naturals} \beta^{'}_n(B_{\rho(\varepsilon)}^c) \leq \varepsilon/3$, where $B_{\rho} := \{\xb \in \reals^l : |\xb| \leq \rho\}$. Then, putting $\delta_n := D_l(\beta_n; \beta^{'}_n)$, fix $n_{\varepsilon} \in \naturals$ for which $\delta_n \leq \min\{\varepsilon/3, \rho(\varepsilon)/2\}$ for every $n \geq n_{\varepsilon}$. Since $\beta_n(A) \leq \beta^{'}_n(A^{\delta_n}) + \delta_n$ holds for every $A \in \mathscr{B}(\reals^l)$, one gets
\[
\beta_n(B_{3\rho(\varepsilon)/2}^c) \leq \beta^{'}_n\big((B_{3\rho(\varepsilon)/2}^c)^{\delta_n}\big) + \delta_n \leq \beta^{'}_n(B_{\rho(\varepsilon)}^c) + \varepsilon/3 \leq 2\varepsilon/3
\]
for every $n \geq n_{\varepsilon}$. The proof is now completed since it is always possible to find a positive number $r(\varepsilon) > 0$ such that $\sup_{n \in \{1, \dots, n_{\varepsilon}-1\}} \beta_n(B_{r(\varepsilon)}^c) \leq \varepsilon/3$.\ \ $\square$ \\


For the statement of the second lemma, let $\big(\Yb_{\nb}^{(1)}, \dots,  \Yb_{\nb}^{(M)}\big)$ and $\big(\Qbb_{\nb}^{(1)}, \dots, \Qbb_{\nb}^{(M)}\big)$ be two families of random elements, indexed by $\nb := (n_1, \dots, n_P) \in \naturals^P$, such that $\Yb_{\nb}^{(i)}$ belongs to $\rd$ and $\Qbb_{\nb}^{(i)}$ is an element of the space of $d\times d$ matrices with real entries. It is also required that 
$\Yb_{\nb}^{(i)}$ and $\Qbb_{\nb}^{(i)}$ depend on $\nb$ only through $n_i + \dots + n_{i+G-1}$, for any $i = 1, \dots, M$. Let $\xi_{\nb}^{(dM)}$ stand for the probability laws of the vector 
$\big(\Yb_{\nb}^{(1)}, \dots,  \Yb_{\nb}^{(M)}\big)$ and assume that
\begin{equation}\label{QconvP}
\big(\Qbb_{\nb}^{(1)}, \dots, \Qbb_{\nb}^{(M)}\big) \rightarrow \big(\overline{\Qbb}^{(1)}, \dots, \overline{\Qbb}^{(M)}\big)
\end{equation}
in probability as $\ n_1, \dots, n_P \rightarrow +\infty$, for suitable non-random $d\times d$ matrices $\overline{\Qbb}^{(1)}, \dots, \overline{\Qbb}^{(M)}$. For completeness, the distance between the two vectors of matrices is measured by $\big(\sum_{i=1}^M \|\Qbb_{\nb}^{(i)} - \overline{\Qbb}^{(i)}\|_F^2\big)^{1/2}$, where $\|\cdot\|_F$ denotes the Frobenius norm.
Moreover, for any elements $\Qbb^{(1)}, \dots, \Qbb^{(M)}$ of the space of $d\times d$ matrices with real entries, write $\mathcal{L}[\Qbb^{(1)}, \dots, \Qbb^{(M)}]$ to indicate the linear mapping $(\xb^{(1)}, \dots, \xb^{(M)}) \ni \reals^{dM} \mapsto \big( \Qbb^{(1)}\xb^{(1)}, \dots, \Qbb^{(M)}\xb^{(M)} \big) \in \reals^{dM}$.
Finally, let $\beta_{\nb}^{(dM)}$ and $\overline{\beta}_{\nb}^{(dM)}$ denote the probability laws of $\mathcal{L}[\Qbb_{\nb}^{(1)}, \dots, \Qbb_{\nb}^{(M)}]\big(\Yb_{\nb}^{(1)}, \dots,  \Yb_{\nb}^{(M)}\big)$ and 
$\mathcal{L}[\overline{\Qbb}^{(1)}, \dots, \overline{\Qbb}^{(M)}]\big(\Yb_{\nb}^{(1)}, \dots,  \Yb_{\nb}^{(M)}\big)$, respectively, and let
$\lambda_{\nb}^{(M)}$ and $\overline{\lambda}_{\nb}^{(M)}$ denote the probability laws of $\big( ^t\Yb_{\nb}^{(1)}\Qbb_{\nb}^{(1)}\Yb_{\nb}^{(1)}, \dots, ^t\Yb_{\nb}^{(M)}\Qbb_{\nb}^{(M)}\Yb_{\nb}^{(M)}\big)$ and \\
$\big( ^t\Yb_{\nb}^{(1)}\overline{\Qbb}^{(1)}\Yb_{\nb}^{(1)}, \dots,  ^t\Yb_{\nb}^{(M)}\overline{\Qbb}^{(M)}\Yb_{\nb}^{(M)}\big)$, respectively.

\begin{lemma} \label{lm:merge}
Let \eqref{QconvP} be in force.
\begin{enumerate}
\item[i)] If $\{\xi_{\nb}^{(dM)}\}_{\nb \in \naturals^P}$ is a tight family of probability laws, there hold 
\begin{itemize}
\item $\big(\Qbb_{\nb}^{(1)}\Yb_{\nb}^{(1)}, \dots, \Qbb_{\nb}^{(M)}\Yb_{\nb}^{(M)}\big) - \big(\overline{\Qbb}^{(1)}\Yb_{\nb}^{(1)}, \dots, \overline{\Qbb}^{(M)}\Yb_{\nb}^{(M)}\big) \rightarrow \mathbf{0}$ 
\item $\big( ^t\Yb_{\nb}^{(1)}\Qbb_{\nb}^{(1)}\Yb_{\nb}^{(1)}, \dots, ^t\!\Yb_{\nb}^{(M)}\Qbb_{\nb}^{(M)}\Yb_{\nb}^{(M)}\big) - \big( ^t\Yb_{\nb}^{(1)}\overline{\Qbb}^{(1)}\Yb_{\nb}^{(1)}, \dots,  ^t\!\Yb_{\nb}^{(M)}\overline{\Qbb}^{(M)}\Yb_{\nb}^{(M)}\big) \rightarrow \mathbf{0}$ 
\end{itemize}
in probability as $\ n_1, \dots, n_P \rightarrow +\infty$. In particular, 
\begin{itemize}
\item $D_{dM}\big(\beta_{\nb}^{(dM)}; \overline{\beta}_{\nb}^{(dM)}\big) \rightarrow 0$
\item $D_M\big(\lambda_{\nb}^{(M)}; \overline{\lambda}_{\nb}^{(M)}\big) \rightarrow 0$ 
\end{itemize}
as $\ n_1, \dots, n_P \rightarrow +\infty$.
\item[ii)] If $\overline{\Qbb}^{(1)}, \dots, \overline{\Qbb}^{(M)}$ are non-singular and $D_{dM}\big(\beta_{\nb}^{(dM)}; \omega_{\nb}^{(dM)}\big) \rightarrow 0$ as $\ n_1, \dots, n_P \rightarrow +\infty$, for some tight family of probability laws 
$\{\omega_{\nb}^{(dM)}\}_{\nb \in \naturals^P}$ on $(\reals^{dM}, \mathscr{B}(\reals^{dM}))$, then $D_{dM}\big(\xi_{\nb}^{(dM)}; \omega_{\nb}^{(dM)} \circ \mathcal{L}[\overline{\Qbb}^{(1)}, \dots, \overline{\Qbb}^{(M)}]\big) \rightarrow 0$ as 
$\ n_1, \dots, n_P \rightarrow +\infty$, where $\circ$ designates the composition of mappings. 
\end{enumerate}
\end{lemma}

\noindent \textbf{Proof of Lemma \ref{lm:merge}}.
$i)$ Thanks to the tightness of $\{\xi_{\nb}^{(dM)}\}_{\nb \in \naturals^P}$, for any $\delta > 0$, there exists a compact subsets of $\rd$, say $K_{\delta}$, such that $\sup_{\nb \in \naturals^P} 
\sup_{i=1, \dots, M} \pp[\Yb_{\nb}^{(i)} \not\in K_{\delta}] \leq \delta$. Whence, for any $\varepsilon > 0$,
\begin{eqnarray*}
\pp\big[\big| \big( \Qbb_{\nb}^{(i)} - \overline{\Qbb}^{(i)} \big) \Yb_{\nb}^{(i)} \big| > \varepsilon \big] &\leq& \pp\big[ \| \Qbb_{\nb}^{(i)} - \overline{\Qbb}^{(i)} \|_F \cdot |\Yb_{\nb}^{(i)}| > \varepsilon \big] \\
&\leq& \pp[\Yb_{\nb}^{(i)} \not\in K_{\delta}] + \pp\big[ \| \Qbb_{\nb}^{(i)} - \overline{\Qbb}^{(i)} \|_F > \varepsilon/(\sup_{\ub \in K_{\delta}} |\ub|) \big] 
\end{eqnarray*}
leading to $\lim_{n_1, \dots, n_P \rightarrow \infty} \pp\big[\big| \big( \Qbb_{\nb}^{(i)} - \overline{\Qbb}^{(i)} \big) \Yb_{\nb}^{(i)} \big| > \varepsilon \big] = 0$, by the arbitrariness of $\delta > 0$. The thesis follows by recalling that the convergence in probability to zero of a sequence of random vectors amounts to the convergence in probability to zero of the sequences of the single components. Moreover, the same argument can be applied to prove the convergence in probability to zero of $\big( ^t\Yb_{\nb}^{(1)}\Qbb_{\nb}^{(1)}\Yb_{\nb}^{(1)}, \dots, ^t\!\Yb_{\nb}^{(M)}\Qbb_{\nb}^{(M)}\Yb_{\nb}^{(M)}\big) - \big( ^t\Yb_{\nb}^{(1)}\overline{\Qbb}^{(1)}\Yb_{\nb}^{(1)}, \dots,  ^t\!\Yb_{\nb}^{(M)}\overline{\Qbb}^{(M)}\Yb_{\nb}^{(M)}\big)$. To prove the merging of the probability distributions, consider the so-called Fortet-Mourier distance, defined as follows. Given two probability measures $\mu_1$ and $\mu_2$ on $(\reals^l, \mathscr{B}(\reals^l))$, set
$$
D_l^{\ast}(\mu_1; \mu_2) := \sup_{h \in \mathcal{BL}_1(\reals^l)} \Big| \int_{\reals^l} h(\ub)\mu_1(\ud\ub) - \int_{\reals^l} h(\ub)  \mu_2(\ud\ub)\Big|
$$ 
where $\mathcal{BL}_1(\reals^l)$ denotes the space of real-valued functions on $\reals^l$ with $\sup_{\ub\in\reals^l} |h(\ub)| + \sup_{\ub\neq\vb} |h(\ub) - h(\vb)|/|\ub - \vb| \leq 1$. To prove that $D_{dM}\big(\beta_{\nb}^{(dM)}; \overline{\beta}_{\nb}^{(dM)}\big) \rightarrow 0$, fix $h \in \mathcal{BL}_1(\reals^{dM})$ and write, for arbitrary $\delta, \eta > 0$,
\begin{gather}
\Big| \int_{\reals^{dM}} h(\ub)\beta_{\nb}^{(dM)}(\ud\ub) - \int_{\reals^{dM}} h(\ub)  \overline{\beta}_{\nb}^{(dM)}(\ud\ub)\Big| \nonumber \\
\leq 2 \sum_{i=1}^M \pp[\Yb_{\nb}^{(i)} \not\in K_{\delta}] + 2 \sum_{i=1}^M \pp[\| \Qbb_{\nb}^{(i)} - \overline{\Qbb}^{(i)}\|_F \geq \eta] + \eta M \sup_{\ub\in K_{\delta}} |\ub| \ . \nonumber 
\end{gather}
Therefore, for any $\varepsilon > 0$, choose $\delta = \varepsilon/(4M)$ and $\eta = \frac{\varepsilon}{2M\sup_{\ub\in K_{\delta}} |\ub|}$ to obtain
$$ 
\limsup_{\ n_1, \dots, n_P \rightarrow +\infty} D_{dM}^{\ast}\big(\beta_{\nb}^{(dM)}; \overline{\beta}_{\nb}^{(dM)}\big) \leq \varepsilon \ , 
$$
which is tantamount to saying that $D_{dM}^{\ast}\big(\beta_{\nb}^{(dM)}; \overline{\beta}_{\nb}^{(dM)}\big) \rightarrow 0$, as $\ n_1, \dots, n_P \rightarrow +\infty$. Finally, the thesis follows from the metric equivalence between the Prokhorov and the Fortet-Mourier distance, stated, e.g., in Theorem 11.3.3 of \cite{Du(01)}. Again, an analogous argument shows that $D_M\big(\lambda_{\nb}^{(M)}; \overline{\lambda}_{\nb}^{(M)}\big) \rightarrow 0$ as $n_1, \dots, n_P \rightarrow \infty$, completing the proof of point $i)$.

To prove point $ii)$, consider again the Fortet-Mourier distance and set $\overline{\mathbb{P}}^{(i)} := \big(\overline{\Qbb}^{(i)}\big)^{-1}$ to write
\begin{eqnarray*}
&& D_{dM}^{\ast}\big(\xi_{\nb}^{(dM)}; \omega_{\nb}^{(dM)} \circ \mathcal{L}[\overline{\Qbb}^{(1)}, \dots, \overline{\Qbb}^{(M)}]\big) \\
&\leq& \sup_{h \in \mathcal{BL}_1(\reals^{dM})} \Big| \ee[h\big(\Yb_{\nb}^{(1)}, \dots,  \Yb_{\nb}^{(M)}\big)] -  \ee[h\big(\overline{\mathbb{P}}^{(1)} \Qbb_{\nb}^{(1)} \Yb_{\nb}^{(1)}, \dots, \overline{\mathbb{P}}^{(M)} \Qbb_{\nb}^{(M)} \Yb_{\nb}^{(M)}\big)]  \Big| \\
&+& \!\!\!\!\! \sup_{h \in \mathcal{BL}_1(\reals^{dM})} \!\! \Big| \ee[h\big(\overline{\mathbb{P}}^{(1)} \Qbb_{\nb}^{(1)} \Yb_{\nb}^{(1)}, \dots, \overline{\mathbb{P}}^{(M)} \Qbb_{\nb}^{(M)} \Yb_{\nb}^{(M)}\big)] - \!\!\int_{\reals^{dM}} \!\!\!\!\! h(\ub)\omega_{\nb}^{(dM)} \!\! \circ \mathcal{L}[\overline{\Qbb}^{(1)}, \dots, \overline{\Qbb}^{(M)}](\ud\ub)  \Big|\ .
\end{eqnarray*}
At this stage, thanks to the properties of the Fortet-Mourier distance, the two summands on the above right-hand side are bounded by $[1+ \max_{i=1, \dots, M} \mathrm{Lip}(\overline{\mathbb{P}}^{(i)})] D_{dM}^{\ast}\big(\beta_{\nb}^{(dM)}; \overline{\beta}_{\nb}^{(dM)}\big)$ and\\ $[1+ \max_{i=1, \dots, M} \mathrm{Lip}(\overline{\mathbb{P}}^{(i)})] D_{dM}^{\ast}\big(\beta_{\nb}^{(dM)}; \omega_{\nb}^{(dM)}\big)$, respectively. Therefore, exploiting once again the metric equivalence between the Prokhorov and the Fortet-Mourier distance, the proof of point $ii)$ follows from point $i)$ after showing that $\{\xi_{\nb}^{(dM)}\}_{\nb \in \naturals^P}$ is a tight family of probability laws. The validity of this last claim can be checked by first invoking Lemma \ref{lm:tight}, which entails the tightness of the family $\{\beta_{\nb}^{(dM)}\}_{\nb \in \naturals^P}$. Here, it is important to stress that $\Yb_{\nb}^{(i)}$ and $\Qbb_{\nb}^{(i)}$ depend on $\nb$ only through $n_i + \dots + n_{i+G-1}$, for any $i = 1, \dots, M$. Finally, combine the well-known Prokhorov and Slutsky theorems to deduce, by means of \eqref{QconvP}, the tightness of $\{\xi_{\nb}^{(dM)}\}_{\nb \in \naturals^P}$ by that of $\{\beta_{\nb}^{(dM)}\}_{\nb \in \naturals^P}$. \ \ $\square$ \\

To introduce the last lemma, start by partitioning the matrix $\mathrm{I}(\thetab_0)$ as follows
$$
\mathrm{I}(\thetab_0) = \left(\begin{array}{c|c}
\mathrm{G}_1(\thetab_0) & \mathrm{G}_2(\thetab_0) \\
\hline
^t\mathrm{G}_2(\thetab_0) & \mathrm{G}_3(\thetab_0)
\end{array} \right)\ ,
$$
where $\mathrm{G}_1(\thetab_0)$, $\mathrm{G}_2(\thetab_0)$ and $\mathrm{G}_3(\thetab_0)$ are sub-matrices of dimension $r\times r$, $r\times (d-r)$ and $(d-r) \times (d-r)$, respectively. Assumption  
C5) entails, in particular, that $\mathrm{G}_3(\thetab_0)$ is symmetric and non-singular, allowing the possibility to introduce the new symmetric matrix
\begin{equation} \label{eq:Htheta0}
\mathrm{H}(\thetab_0) := \left(\begin{array}{c|c}
\mathbf{0} & \mathbf{0} \\
\hline
\mathbf{0} & \mathrm{G}_3(\thetab_0)^{-1}
\end{array} \right)\ .
\end{equation}
With this notation, one can provide an alternative representation for the probability distribution $\chi^2_{M,r}(\mathrm{R}_M)$, stated in the following 
\begin{lemma} \label{lm:chi}
If $(^t\mathbf{G}_1, \dots, ^t\mathbf{G}_M)$ is a Gaussian $dM$-dimensional (column) random vector with zero means and covariance matrix 
\begin{equation}\label{eq:covrho}
\left(\begin{array}{c|c|c|c}
\rho_{1,1} \mathrm{I}(\thetab_0) & \rho_{1,2} \mathrm{I}(\thetab_0) & \ldots & \rho_{1,M} \mathrm{I}(\thetab_0) \\
\hline
\rho_{2,1} \mathrm{I}(\thetab_0) & \rho_{2,2} \mathrm{I}(\thetab_0) & \ldots & \rho_{2,M} \mathrm{I}(\thetab_0) \\
\hline
\vdots & \vdots & \ddots & \vdots \\
\hline
\rho_{M,1} \mathrm{I}(\thetab_0) & \rho_{M,2} \mathrm{I}(\thetab_0) & \ldots & \rho_{M,M} \mathrm{I}(\thetab_0)
\end{array} \right) \ ,
\end{equation}
then $\chi^2_{M,r}(\mathrm{R}_M)$ coincides with the probability distribution of the $M$-dimensional random vector whose $i$-th component is equal to
$^t\mathbf{G}_i\ ^t[\mathrm{Id}_{d\times d} - \mathrm{I}(\thetab_0)\mathrm{H}(\thetab_0)]\ \mathrm{I}(\thetab_0)^{-1}\ [\mathrm{Id}_{d\times d} - \mathrm{I}(\thetab_0)\mathrm{H}(\thetab_0)]\ \mathbf{G}_i$, for $i = 1, \dots, M$.
\end{lemma}

\noindent \textbf{Proof of Lemma \ref{lm:chi}}.
By the definition of $\mathrm{H}(\thetab_0)$, one immediately gets $\mathrm{H}(\thetab_0)\mathrm{I}(\thetab_0)\mathrm{H}(\thetab_0) = \mathrm{H}(\thetab_0)$, yielding that
$^t[\mathrm{Id}_{d\times d} - \mathrm{I}(\thetab_0)\mathrm{H}(\thetab_0)] \mathrm{I}(\thetab_0)^{-1} [\mathrm{Id}_{d\times d} - \mathrm{I}(\thetab_0)\mathrm{H}(\thetab_0)] = \mathrm{I}(\thetab_0)^{-1} - \mathrm{H}(\thetab_0)$. 
Then, introduce the $dM$-dimensional random vector $(^t\mathbf{Z}_1, \dots, ^t\mathbf{Z}_M)$ by putting $\mathbf{Z}_i = (Z_{1,i}, \dots, Z_{d,i}) := \mathrm{I}(\thetab_0)^{-1/2} \mathbf{G}_i$, 
whose probability distribution is Gaussian with zero means and covariance matrix equal to
$$
\left(\begin{array}{c|c|c|c}
\rho_{1,1} \mathrm{Id}_{d \times d} & \rho_{1,2} \mathrm{Id}_{d \times d} & \ldots & \rho_{1,M} \mathrm{Id}_{d \times d} \\
\hline
\rho_{2,1} \mathrm{Id}_{d \times d} & \rho_{2,2} \mathrm{Id}_{d \times d} & \ldots & \rho_{2,M} \mathrm{Id}_{d \times d} \\
\hline
\vdots & \vdots & \ddots & \vdots \\
\hline
\rho_{M,1} \mathrm{Id}_{d \times d} & \rho_{M,2} \mathrm{Id}_{d \times d} & \ldots & \rho_{M,M} \mathrm{Id}_{d \times d}
\end{array} \right) \ .
$$
To conclude, rewrite the definition of $\chi^2_{M,r}(\mathrm{R}_M)$ as probability law of the random vector $(|\mathrm{P}_{d,r} \mathbf{Z}_1|^2, \dots, |\mathrm{P}_{d,r} \mathbf{Z}_M|^2)$, with 
$$
\mathrm{P}_{d,r} := \left(\begin{array}{c|c}
\mathrm{Id}_{r \times r} & \mathbf{0} \\
\hline
\mathbf{0} & \mathbf{0}
\end{array} \right)\ ,
$$
and notice that $\mathrm{I}(\thetab_0)^{1/2}\ [\mathrm{I}(\thetab_0)^{-1} - \mathrm{H}(\thetab_0)]\ \mathrm{I}(\thetab_0)^{1/2} = \mathrm{P}_{d,r}$, yielding
\begin{eqnarray}
|\mathrm{P}_{d,r} \mathbf{Z}_i|^2 &=& ^t\mathbf{Z}_i ^t\mathrm{P}_{d,r} \mathrm{P}_{d,r} \mathbf{Z}_i =\ ^t\mathbf{Z}_i \mathrm{P}_{d,r} \mathbf{Z}_i \nonumber \\
&=&\ ^t\mathbf{G}_i\ \mathrm{I}(\thetab_0)^{-1/2} \mathrm{I}(\thetab_0)^{1/2} [\mathrm{I}(\thetab_0)^{-1} - \mathrm{H}(\thetab_0)] \mathrm{I}(\thetab_0)^{1/2} \mathrm{I}(\thetab_0)^{-1/2} \mathbf{G}_i \nonumber \\
&=&\  ^t\mathbf{G}_i\ ^t[\mathrm{Id}_{d\times d} - \mathrm{I}(\thetab_0)\mathrm{H}(\thetab_0)] \mathrm{I}(\thetab_0)^{-1} [\mathrm{Id}_{d\times d} - \mathrm{I}(\thetab_0)\mathrm{H}(\thetab_0)] \mathbf{G}_i\ . \nonumber
\end{eqnarray}
for $i = 1, \dots, M$. \ \ $\square$ \\

The way is now paved for the proof of the main theorems. \\

\noindent \textbf{Proof of Theorem \ref{thm:cramer}}
At the beginning, introduce the symbols 
$$
L_{n_i, \dots, n_{i + G - 1}}(\thetab; \mathbf{X}^{(i)}; \dots; \mathbf{X}^{(i + G - 1)}) := \prod_{p = i}^{i + G - 1} \prod_{j = 1}^{n_p} f(X_j^{(p)}\ |\ \thetab)
$$ 
and 
\begin{align*}
\ell^{'}_{n_i, \dots, n_{i + G - 1}}(\thetab) &:= \nabla_{\thetab} \log [L_{n_i, \dots, n_{i + G - 1}}(\thetab; \mathbf{X}^{(i)}; \dots; \mathbf{X}^{(i + G - 1)})] \\
&= \sum_{p = i}^{i + G - 1} \sum_{j = 1}^{n_p} \nabla_{\thetab} \log[f(X_j^{(p)}\ |\ \thetab)]
\end{align*}
for $i = 1, \dots, M$. Thanks to C7), the MLE's $\hat{\thetab}_{n_i, \dots, n_{i + G - 1}} = \tb_{n_i + \dots + n_{i + G - 1}}(\mathbf{X}^{(i)}; \dots; \mathbf{X}^{(i + G - 1)})$ relative to 
$L_{n_i, \dots, n_{i + G - 1}}(\cdot; \mathbf{X}^{(i)}; \dots; \mathbf{X}^{(i + G - 1)})$, being internal points of $\Theta$, satisfy $\ell^{'}_{n_i, \dots, n_{i + G - 1}}(\hat{\thetab}_{n_i, \dots, n_{i + G - 1}}) = \mathbf{0}$
for $i = 1, \dots, M$. It is also well-known that, under the assumptions C1)-C7), the MLE's are strongly consistent, in the sense that $\hat{\thetab}_{n_i, \dots, n_{i + G - 1}} \rightarrow \thetab_0$ almost surely as $n_1, \dots, n_P \rightarrow +\infty$, whenever the common density of the $X^{(p)}_j$'s is $f(\cdot;\thetab_0)$. For a proof, see Chapters 17-18 of \cite{Fer(02)}. Then, the Taylor formula with integral remainder entails
\begin{equation} \label{Taylor11}
\frac{1}{\sqrt{\sum_{k = i}^{i + G - 1} n_k}} \ell^{'}_{n_i, \dots, n_{i + G - 1}}(\thetab_0) = \sqrt{\sum_{k = i}^{i + G - 1} n_k} \cdot \mathrm{B}_{n_i, \dots, n_{i + G - 1}} (\hat{\thetab}_{n_i, \dots, n_{i + G - 1}} - \thetab_0)
\end{equation}
where 
$$
\mathrm{B}_{n_i, \dots, n_{i + G - 1}} := - \int_0^1 \frac{1}{\sum_{k = i}^{i + G - 1} n_k} \Big\{\sum_{p = i}^{i + G - 1} \sum_{j = 1}^{n_p} \mathrm{M}(X_j^{(p)}; \thetab_0 + u(\hat{\thetab}_{n_i, \dots, n_{i + G - 1}} - \thetab_0))\Big\} \ud u
$$
and $\mathrm{M}(x; \mathbf{t})$ is the $d \times d$ matrix given by $\left(\frac{\partial^2}{\partial t_k \partial t_h} \log f(x\ |\ \mathbf{t})\right)_{k, h = 1, \dots, d}$. It is well-known that $\mathrm{B}_{n_i, \dots, n_{i + G - 1}} \rightarrow \mathrm{I}(\thetab_0)$ almost surely as $n_1, \dots, n_P \rightarrow +\infty$, meaning that $\|\mathrm{B}_{n_i, \dots, n_{i + G - 1}} - \mathrm{I}(\thetab_0)\|_F \rightarrow 0$ almost surely.
See, e.g., the final part of the proof of Theorem 18 in \cite{Fer(02)}. Therefore, the original problem is traced back to the approximation of the sequence $\{\zeta_{n_1, \dots, n_P}^{(dM)}\}_{n_1, \dots, n_P \geq 1}$,
the single $\zeta_{n_1, \dots, n_P}^{(dM)}$ being the probability distribution of the $Md$-dimensional random vector
$$
\mathbf{U}_{\nb} := \Big(\frac{1}{\sqrt{\sum_{k = 1}^G n_k}} \ell^{'}_{n_1, \dots, n_G}(\thetab_0), \dots, \frac{1}{\sqrt{\sum_{k = M}^P n_k}} \ell^{'}_{n_M, \dots, n_P}(\thetab_0)\Big)
$$
where, by definition,
$$
\frac{1}{\sqrt{\sum_{k = i}^{i+G-1} n_k}} \ell^{'}_{n_i, \dots, n_{i+G-1}}(\thetab_0) =
\sqrt{\sum_{k = i}^{i+G-1} n_k} \Big(\frac{1}{\sum_{k = i}^{i+G-1} n_k} \sum_{p = i}^{i+G-1} \sum_{j = 1}^{n_p} \mathbf{\Psi}(X_j^{(p)}; \thetab_0)\Big)
$$
for $i = 1, \dots, M$, with $\mathbf{\Psi}(x; \mathbf{t}) := \nabla_{\thetab} \log[f(x | \thetab)]_{\big| \thetab = \tb}$. The random vectors 
$\{\mathbf{\Psi}(X_j^{(p)}; \thetab_0)\}_{\substack{j = 1, \dots, n_p \\ p = 1, \dots, P}}$ are i.i.d. and, from C3), it follows that
\begin{eqnarray}
\ee_{\thetab_0}[\mathbf{\Psi}(X_j^{(p)}; \thetab_0)] &=& \mathbf{0} \label{eq:mediaPsi} \\
\textsf{Cov}_{\thetab_0}(\Psi^{(k)}(X_j^{(p)}; \thetab_0), \Psi^{(h)}(X_j^{(p)}; \thetab_0)) &=& \mathrm{I}_{k, h}(\thetab_0) \label{eq:varianzaPsi}
\end{eqnarray}
where $\Psi^{(k)}(X_j^{(p)}; \thetab_0)$ denotes the $k$-th coordinate of $\mathbf{\Psi}(X_j^{(p)}; \thetab_0)$. Introduce the independent $d$-dimensional random vectors $\mathbf{S}_p := \sum_{j = 1}^{n_p} \mathbf{\Psi}(X_j^{(p)}; \thetab_0)$ for $p = 1, \dots, P$ and consider the characteristic function of $\mathbf{U}_{\nb}$, given by
\begin{align*}
\hat{\zeta}_{n_1, \dots, n_P}^{(dM)}(\xib_1, \dots, \xib_M) &= \ee_{\thetab_0}\Big[\exp\Big\{\sum_{m=1}^M \frac{i\xib_m \bullet \sum_{p=m}^{m+G-1} \mathbf{S}_p}{\sqrt{\sum_{k = m}^{m+G-1} n_k}} \Big\}\Big] \\
&= \ee_{\thetab_0}\Big[\exp\Big\{\sum_{p=1}^P \mathbf{S}_p \bullet \sum_{\substack{m = 1, \dots, M \\ m \leq p \leq m+G-1}}
\frac{i\xib_m}{\sqrt{\sum_{k = m}^{m+G-1} n_k}}\Big\}\Big] \\
&= \prod_{p=1}^P \varphi^{n_p}\Big( \sum_{\substack{m = 1, \dots, M \\ m \leq p \leq m+G-1}}
\frac{\xib_m}{\sqrt{\sum_{k = m}^{m+G-1} n_k}}\Big)
\end{align*}
where $\bullet$ stands for the standard scalar product in $\rd$ and $\varphi(\xib) := \ee_{\thetab_0}[\exp\{i \xib \bullet \mathbf{\Psi}(X_j^{(p)}; \thetab_0)\}]$, with $\xib \in \rd$. At this stage, notice that 
\begin{eqnarray}
&& \sum_{p=1}^P n_p \sum_{\substack{m = 1, \dots, M \\ m \leq p \leq m+G-1}} \sum_{\substack{l = 1, \dots, M \\ l \leq p \leq l+G-1}}\Big(\frac{1}{\sqrt{\sum_{k = m}^{m+G-1} \sum_{h = l}^{l+G-1} n_k n_h}}\Big)\ ^t\xib_m \mathrm{I}(\theta_0)\xib_l \nonumber \\
&=& \sum_{\substack{m, l = 1, \dots, M \\ |m-l| < G}} \Big(\frac{\sum_{p = a(l,m)}^{b(l, m)}n_p}{\sqrt{\sum_{k = m}^{m+G-1} \sum_{h = l}^{l+G-1} n_k n_h}}\Big)\ ^t\xib_m \mathrm{I}(\theta_0)\xib_l \nonumber
\end{eqnarray}
holds with $a(l,m) := \max\{l, m\}$ and $b(l,m) := \min\{l, m\} + G - 1$. Therefore, after recalling \eqref{eq:rhoij}, the above quadratic form proves to be equal to
$$
\left(\begin{array}{cccc}
\xib_1 & \xib_2 & \ldots & \xib_M
\end{array} \right)
\left(\begin{array}{c|c|c|c}
\rho_{1,1} \mathrm{I}(\thetab_0) & \rho_{1,2} \mathrm{I}(\thetab_0) & \ldots & \rho_{1,M} \mathrm{I}(\thetab_0) \\
\hline
\rho_{2,1} \mathrm{I}(\thetab_0) & \rho_{2,2} \mathrm{I}(\thetab_0) & \ldots & \rho_{2,M} \mathrm{I}(\thetab_0) \\
\hline
\vdots & \vdots & \ddots & \vdots \\
\hline
\rho_{M,1} \mathrm{I}(\thetab_0) & \rho_{M,2} \mathrm{I}(\thetab_0) & \ldots & \rho_{M,M} \mathrm{I}(\thetab_0)
\end{array} \right)
\left(\begin{array}{c}
\xib_1 \\
\xib_2 \\
\vdots \\
\xib_M
\end{array} \right)\ .
$$
These remarks are conducive to the introduction of the metric 
$$
D_{gtw}(\mu_1; \mu_2) := \sup_{\xib \in \reals^l\setminus{\{\mathbf{0}\}}} \frac{|\hat{\mu}_1(\xib) - \hat{\mu}_2(\xib)|}{|\xib|^2}
$$
with $\hat{\mu}_j(\xib) := \int_{\reals^l} \exp\{i \xib \bullet \xb\} \mu_j(\ud \xb)$, $j=1,2$, which is defined for any pair $(\mu_1, \mu_2)$ of probability measures on $(\reals^l, \mathscr{B}(\reals^l))$ such that $\int_{\reals^l} |\xb|^2 \mu_1(\ud \xb) + \int_{\reals^l} |\xb|^2 \mu_2(\ud \xb) < +\infty$ and $\int_{\reals^l} \xb \mu_1(\ud \xb) = \int_{\reals^l} \xb \mu_2(\ud \xb)$. It is well-known (see, e.g., Section 5 of \cite{Gab(95)})
that there exists a modulus of continuity $\omega_l : [0,+\infty) \rightarrow [0,+\infty)$ such that, for any such pair $(\mu_1, \mu_2)$, there holds $D_l(\mu_1; \mu_2) \leq \omega_l(D_{gtw}(\mu_1; \mu_2))$. 

The main step consists now in the proof of
\begin{equation} \label{cramerquasi}
D_l\big(\zeta_{n_1, \dots, n_P}^{(dM)}; \gamma^{(dM)}(\mathrm{R}_M, \mathrm{I}(\thetab_0)) \big) \rightarrow 0
\end{equation}
as $\ n_1, \dots, n_P \rightarrow +\infty$, $\gamma^{(dM)}(\mathrm{R}_M, \mathrm{I}(\thetab_0))$ denoting the $dM$-dimensional Gaussian distribution with zero means and covariance matrix \eqref{eq:covrho}.
But, in view of the above considerations, it is enough to prove $D_{gtw}\big(\zeta_{n_1, \dots, n_P}^{(dM)}; \gamma^{(dM)}(\mathrm{R}_M, \mathrm{I}(\thetab_0)) \big) \rightarrow 0$ as $\ n_1, \dots, n_P \rightarrow +\infty$. To prove this last claim, recall that the matrix $\mathrm{R}_M$ is positive-definite and invoke Lemma 1 in Section 27 of \cite{Bil(95)} to obtain
\begin{eqnarray*}
&& D_{gtw}\big(\zeta_{n_1, \dots, n_P}^{(dM)}; \gamma^{(dM)}(\mathrm{R}_M, \mathrm{I}(\thetab_0)) \big) \leq \sum_{p=1}^P \sup_{\xib \in \reals^{dM}\setminus{\{\mathbf{0}\}}} \!\!\!\! |\xib|^{-2} \!\cdot \Big| \varphi^{n_p}\Big( \sum_{\substack{m = 1, \dots, M \\ m \leq p \leq m+G-1}}\frac{\xib_m}{\sqrt{\sum_{k = m}^{m+G-1} n_k}}\Big) \\
&-& \exp\Big\{-\frac12  \sum_{\substack{m = 1, \dots, M \\ m \leq p \leq m+G-1}}  \sum_{\substack{l = 1, \dots, M \\ l \leq p \leq l+G-1}} \Big(\frac{n_p}{\sqrt{\sum_{k = m}^{m+G-1} \sum_{h = l}^{l+G-1} n_k n_h}}\Big)\ ^t\xib_m \mathrm{I}(\theta_0)\xib_l \Big\}\Big|
\end{eqnarray*}
where $\xib = (\xib_1, \dots, \xib_M)$. For fixed $\varepsilon > 0$, choose $T_{\varepsilon} > 0$ large enough to guarantee $T_{\varepsilon}^{-2} \leq \varepsilon/(2P)$. Now, thanks to 
\eqref{eq:mediaPsi}-\eqref{eq:varianzaPsi}, the usual properties of characteristic functions (see, e.g. Sections 26 and 29 of \cite{Bil(95)}) show that, for any $p \in \{1, \dots, P\}$,
\begin{eqnarray*}
&& \varphi\Big( \sum_{\substack{m = 1, \dots, M \\ m \leq p \leq m+G-1}} \frac{\xib_m}{\sqrt{\sum_{k = m}^{m+G-1} n_k}}\Big) \\
&=& 1 - \frac12 \!\!\!\!\sum_{\substack{m = 1, \dots, M \\ m \leq p \leq m+G-1}}  \sum_{\substack{l = 1, \dots, M \\ l \leq p \leq l+G-1}} \Big(\frac{1}{\sqrt{\sum_{k = m}^{m+G-1} \sum_{h = l}^{l+G-1} n_k n_h}}\Big)\ ^t\xib_m \mathrm{I}(\theta_0)\xib_l + \mathcal{E}_{n_1, \dots, n_P}^{(p)}(\xib)
\end{eqnarray*}
holds with $\sup_{\ 0 < |\xib| \leq T_{\varepsilon}} |\xib|^{-2} |\mathcal{E}_{n_1, \dots, n_P}^{(p)}(\xib)| = o(1/n_p)$. It is now routine to utilize the usual arguments provided to prove the classical CLT (see, e.g., Section 27 of \cite{Bil(95)}), to conclude that
$$
\limsup_{\ n_1, \dots, n_P \rightarrow +\infty} D_{gtw}\big(\zeta_{n_1, \dots, n_P}^{(dM)}; \gamma^{(dM)}(\mathrm{R}_M, \mathrm{I}(\thetab_0)) \big) \leq \varepsilon
$$
is in force for any $\varepsilon > 0$, yielding \eqref{cramerquasi} in view of the arbitrariness of $\varepsilon$.

The actual proof of \eqref{eq:cramer} follows from point $ii)$ of Lemma \ref{lm:merge}. In fact, thanks to \eqref{Taylor11}, it is possible to put 
\begin{eqnarray*}
\big(\Yb_{\nb}^{(1)}, \dots, \Yb_{\nb}^{(M)}\big) &=& \Big(\sqrt{\sum_{k = 1}^G n_k} \cdot (\hat{\thetab}_{n_1, \dots, n_G} - \thetab_0), \dots, \sqrt{\sum_{k = M}^P n_k} \cdot (\hat{\thetab}_{n_M, \dots, n_P} - \thetab_0)\Big) \\
\big(\Qbb_{\nb}^{(1)}, \dots, \Qbb_{\nb}^{(M)}\big) &=& \big(\mathrm{B}_{n_1, \dots, n_G}, \dots, \mathrm{B}_{n_M, \dots, n_P}\big) \\
\big(\overline{\Qbb}^{(1)}, \dots, \overline{\Qbb}^{(M)}\big) &=& \big(\mathrm{I}(\thetab_0), \dots, \mathrm{I}(\thetab_0)\big) 
\end{eqnarray*}
so that the validity of \eqref{QconvP} is guaranteed, as well as the non-singularity of the $\overline{\Qbb}^{(i)}$'s. In addition, setting $\beta_{\nb}^{(dM)} = \zeta_{n_1, \dots, n_P}^{(dM)}$ and $\omega_{\nb}^{(dM)}
= \gamma^{(dM)}(\mathrm{R}_M, \mathrm{I}(\thetab_0))$ entails that $D_{dM}\big(\beta_{\nb}^{(dM)}; \omega_{\nb}^{(dM)}\big) \rightarrow 0$ as $\ n_1, \dots, n_P \rightarrow +\infty$ amounts to \eqref{cramerquasi}. The tightness of $\{\omega_{\nb}^{(dM)}\}_{\nb \in \naturals^P}$ follows from the boundedness of the $\rho_{i,j}$'s. Finally, the elementary properties of the Gaussian distributions 
lead to the identity $\gamma^{(dM)}(\mathrm{R}_M, \mathrm{I}(\thetab_0)) \circ \mathcal{L}[\mathrm{I}(\thetab_0), \dots, \mathrm{I}(\thetab_0)] = \gamma^{(dM)}(\mathrm{R}_M, \mathrm{I}(\thetab_0)^{-1})$.\ \ $\square$ \\


\noindent \textbf{Proof of Theorem \ref{thm:wilks}}.
Observe that 
$$
\Xi_i = 2[\ell_{n_i, \dots, n_{i+G-1}}(\hat{\thetab}_{n_i, \dots, n_{i + G - 1}}) - \ell_{n_i, \dots, n_{i+G-1}}(\thetab_{n_i, \dots, n_{i + G - 1}}^{\ast})]
$$ 
for $i = 1, \dots, M$, where $\hat{\thetab}_{n_i, \dots, n_{i + G - 1}}$ ($\thetab_{n_i, \dots, n_{i + G - 1}}^{\ast}$, respectively) stands for the MLE over $\Theta$ ($\Theta_0$, respectively), based on the sample $(\mathbf{X}^{(i)}; \dots; \mathbf{X}^{(i+G-1)})$. Without loss of generality, assume that $\thetab_0$ coincides with the origin and that
\begin{equation} \label{eq:defH0}
\Theta_0 = \{\thetab = (\theta^{(1)}, \dots, \theta^{(d)}) \in \Theta\ |\ \theta^{(1)} =  \dots = \theta^{(r)} = 0\} 
\end{equation}
since, by definition of sub-manifold, it is always possible to reduce the problem to this situation after a local change of coordinates. Then, the proof is split into two parts.

As for the former part, the first step consists in the introduction of the following three $dM$-dimensional random vectors, indexed by $\nb := (n_1, \dots, n_P) \in \naturals^P$:
\begin{itemize}
\item $\mathbf{U}_{\nb} := \big(\mathbf{U}_{\nb}^{(1)}, \dots, \mathbf{U}_{\nb}^{(M)}\big)$ with $\mathbf{U}_{\nb}^{(i)} := (n_i + \dots + n_{i+G-1})^{-1/2} \ell^{'}_{n_i, \dots, n_{i+G-1}}(\thetab_0)$
\item $\mathbf{V}_{\nb} := \big(\mathbf{V}_{\nb}^{(1)}, \dots, \mathbf{V}_{\nb}^{(M)}\big)$ with $\mathbf{V}_{\nb}^{(i)} := (n_i + \dots + n_{i+G-1})^{-1/2} \ell^{'}_{n_i, \dots, n_{i+G-1}}(\thetab_{n_i, \dots, n_{i + G - 1}}^{\ast})$
\item $\mathbf{W}_{\nb} := \big(\mathbf{W}_{\nb}^{(1)}, \dots, \mathbf{W}_{\nb}^{(M)}\big)$ with $\mathbf{W}_{\nb}^{(i)} := [\mathrm{Id}_{d\times d} - \mathrm{I}(\thetab_0)\mathrm{H}(\thetab_0)] \mathbf{U}_{\nb}^{(i)}$
\end{itemize}
where $\mathrm{H}(\thetab_0)$ has been defined by (\ref{eq:Htheta0}). An alternative expression for $\mathbf{V}_{\nb}^{(i)}$ can be obtained by a Taylor expansion of $\ell^{'}_{n_i, \dots, n_{i+G-1}}(\thetab_{n_i, \dots, n_{i + G - 1}}^{\ast})$ about $\thetab_0$ with integral remainder, yielding
\begin{equation} \label{eq:Taylor7}
\mathbf{V}_{\nb}^{(i)} = \mathbf{U}_{\nb}^{(i)} + \mathbb{Q}_{n_i + \dots + n_{i + G - 1}}^{\ast} \sqrt{n_i+ \dots + n_{i + G - 1}} \big(\thetab_{n_i, \dots, n_{i + G - 1}}^{\ast} - \thetab_0\big)
\end{equation}
for $i = 1, \dots, M$, with 
$$
\mathbb{Q}_{n_i + \dots + n_{i + G - 1}}^{\ast} := \frac{1}{n_i + \dots + n_{i + G - 1}} \int_0^1 \mathrm{Hess}[\ell_{n_i, \dots, n_{i+G-1}}]\Big(\thetab_0 + t\big(\thetab_{n_i, \dots, n_{i + G - 1}}^{\ast} - \thetab_0\big) \Big) \ud t  \ .
$$
The form of $\Theta_0$ given in (\ref{eq:defH0}) entails that $[\mathrm{Id}_{d\times d} - \mathrm{P}_{d,r}]\ \ell^{'}_{n_i, \dots, n_{i+G-1}}(\thetab_{n_i, \dots, n_{i + G - 1}}^{\ast}) = \mathbf{0}$, implying $\mathrm{H}(\thetab_0)
\ell^{'}_{n_i, \dots, n_{i+G-1}}(\thetab_{n_i, \dots, n_{i + G - 1}}^{\ast}) = \mathbf{0}$. Whence,
$$
\mathrm{H}(\thetab_0) \mathbf{U}_{\nb}^{(i)} + \mathrm{H}(\thetab_0) \mathbb{Q}_{n_i + \dots + n_{i + G - 1}}^{\ast} \sqrt{n_i+ \dots + n_{i + G - 1}}\big(\thetab_{n_i, \dots, n_{i + G - 1}}^{\ast} - \thetab_0\big) = \mathbf{0}
$$
for $i = 1, \dots, M$. Taking account that $\mathrm{H}(\thetab_0)\mathrm{I}(\thetab_0) \big(\thetab_{n_i, \dots, n_{i + G - 1}}^{\ast} - \thetab_0\big) = \big(\thetab_{n_i, \dots, n_{i + G - 1}}^{\ast} - \thetab_0\big)$ and that
$\mathbb{Q}_{n_i + \dots + n_{i + G - 1}}^{\ast} \rightarrow -\mathrm{I}(\thetab_0)$ in probability, one gets
$$
\mathrm{H}(\thetab_0) \mathbf{U}_{\nb}^{(i)} - \mathrm{H}(\thetab_0)\mathrm{I}(\thetab_0) \sqrt{n_i+ \dots + n_{i + G - 1}} \big(\thetab_{n_i, \dots, n_{i + G - 1}}^{\ast} - \thetab_0\big) \rightarrow \mathbf{0}
$$
in probability as $\ n_1, \dots, n_P \rightarrow +\infty$, for $i = 1, \dots, M$. The combination of this fact with (\ref{eq:Taylor7}) shows that
$\mathbf{V}_{\nb} - \mathbf{W}_{\nb} \rightarrow \mathbf{0}$ in probability as $\ n_1, \dots, n_P \rightarrow +\infty$, for $i = 1, \dots, M$. Since the tightness of the family of probability laws of the 
$\mathbf{U}_{\nb}$'s, and hence the tightness of the family of probability laws of the $\mathbf{W}_{\nb}$'s, has been already checked in the proof of Theorem \ref{thm:cramer}, the argument used to prove Lemma
\ref{lm:merge} yields 
\begin{equation} \label{eq:Merging1}
D_{dM}\big(\rho_{n_1, \dots, n_P}^{(dM)}; \sigma_{n_1, \dots, n_P}^{(dM)} \big) \rightarrow 0
\end{equation}
as $\ n_1, \dots, n_P \rightarrow +\infty$, $\rho_{n_1, \dots, n_P}^{(dM)}$ and $\sigma_{n_1, \dots, n_P}^{(dM)}$ standing for the probability distributions of $\mathbf{V}_{\nb}$ and $\mathbf{W}_{\nb}$, respectively. 
Moreover, a combination of the classical CLT with the mapping theorem gives
\begin{equation} \label{eq:Merging2}
D_{dM}\big(\sigma_{n_1, \dots, n_P}^{(dM)}; \tau_{n_1, \dots, n_P}^{(dM)} \big) \rightarrow 0
\end{equation}
as $\ n_1, \dots, n_P \rightarrow +\infty$, where $\tau_{n_1, \dots, n_P}^{(dM)}$ denotes the probability distribution of the random vector $\big([\mathrm{Id}_{d\times d} - \mathrm{I}(\thetab_0)\mathrm{H}(\thetab_0)] \mathbf{G}_1, \dots, [\mathrm{Id}_{d\times d} - \mathrm{I}(\thetab_0)\mathrm{H}(\thetab_0)] \mathbf{G}_M\big)$, $(\mathbf{G}_1, \dots, \mathbf{G}_M)$ being the same as in Lemma \ref{lm:chi}. 


In the latter part of the proof, taking account that $\ell^{'}_{n_i, \dots, n_{i+G-1}}(\hat{\thetab}_{n_i, \dots, n_{i + G - 1}}) = \mathbf{0}$ for all large $n_i, \dots, n_{i+G-1}$, expand 
$\ell_{n_i, \dots, n_{i+G-1}}(\thetab_{n_i, \dots, n_{i + G - 1}}^{\ast})$ by Taylor's formula about $\hat{\thetab}_{n_i, \dots, n_{i + G - 1}}$, to obtain
\begin{eqnarray}
&& \ell_{n_i, \dots, n_{i+G-1}}(\thetab_{n_i, \dots, n_{i + G - 1}}^{\ast}) = \ell_{n_i, \dots, n_{i+G-1}}(\hat{\thetab}_{n_i, \dots, n_{i + G - 1}}) - (n_i + \dots + n_{i + G - 1}) \times \nonumber \\
&\times&\ ^t\big(\thetab_{n_i, \dots, n_{i + G - 1}}^{\ast\ast} - \hat{\thetab}_{n_i, \dots, n_{i + G - 1}}\big) 
\hat{\mathbb{Q}}_{n_i + \dots + n_{i + G - 1}}^{\ast\ast} \big(\thetab_{n_i, \dots, n_{i + G - 1}}^{\ast} - \hat{\thetab}_{n_i, \dots, n_{i + G - 1}}\big) \nonumber 
\end{eqnarray}
where the matrix $\hat{\mathbb{Q}}_{n_i + \dots + n_{i + G - 1}}^{\ast\ast}$ is given by
\begin{eqnarray}
-\frac{1}{n_i + \dots + n_{i + G - 1}} \int_0^1 \int_0^1 t &\cdot& \mathrm{Hess}[\ell_{n_i, \dots, n_{i+G-1}}]\Big(\hat{\thetab}_{n_i, \dots, n_{i + G - 1}} \nonumber \\
&+& st\big(\thetab_{n_i, \dots, n_{i + G - 1}}^{\ast} - \hat{\thetab}_{n_i, \dots, n_{i + G - 1}}\big) \Big) \ud s \ud t \ . \nonumber 
\end{eqnarray}
by virtue of the integral form of the reminder. Whence,
\begin{eqnarray}
\Xi_i &=& 2\sqrt{n_i + \dots + n_{i + G - 1}}\ ^t\big(\thetab_{n_i, \dots, n_{i + G - 1}}^{\ast} - \hat{\thetab}_{n_i, \dots, n_{i + G - 1}}\big) \hat{\mathbb{Q}}_{n_i + \dots + n_{i + G - 1}}^{\ast\ast} \times
\nonumber \\
&\times& \sqrt{n_i + \dots + n_{i + G - 1}}\ \big(\thetab_{n_i, \dots, n_{i + G - 1}}^{\ast} - \hat{\thetab}_{n_i, \dots, n_{i + G - 1}}\big) \label{eq:start1}
\end{eqnarray}
for $i = 1, \dots, M$ and for all large $n_1, \dots, n_P$. Apropos of the asymptotic behavior of $\sqrt{n_i + \dots + n_{i + G - 1}}\ ^t\big(\thetab_{n_i, \dots, n_{i + G - 1}}^{\ast} - \hat{\thetab}_{n_i, \dots, n_{i + G - 1}}\big)$, expand $\mathbf{V}_{\nb}^{(i)}$  by Taylor's formula about $\hat{\thetab}_{n_i, \dots, n_{i + G - 1}}$, to obtain
$$
\mathbf{V}_{\nb}^{(i)} = \hat{\mathbb{Q}}_{n_i + \dots + n_{i + G - 1}}^{\ast}\sqrt{n_i + \dots + n_{i + G - 1}}\  \big(\thetab_{n_i, \dots, n_{i + G - 1}}^{\ast} - \hat{\thetab}_{n_i, \dots, n_{i + G - 1}}\big) 
$$
where the matrix $\hat{\mathbb{Q}}_{n_i + \dots + n_{i + G - 1}}^{\ast}$ is given by
$$
\frac{1}{n_i + \dots + n_{i + G - 1}} \int_0^1 \mathrm{Hess}[\ell_{n_i, \dots, n_{i+G-1}}]\Big(\hat{\thetab}_{n_i, \dots, n_{i + G - 1}} + t\big(\thetab_{n_i, \dots, n_{i + G - 1}}^{\ast} - 
\hat{\thetab}_{n_i, \dots, n_{i + G - 1}}\big) \Big) \ud t
$$
by virtue of the integral form of the reminder. Thanks to the tightness of $\rho_{n_1, \dots, n_P}^{(dM)}$ and the fact that $\hat{\mathbb{Q}}_{n_i + \dots + n_{i + G - 1}}^{\ast} \rightarrow -\mathrm{I}(\thetab_0)$ in probability as $n_i, \dots, n_{i + G - 1} \rightarrow +\infty$ (see, e.g., the proof of Theorem 22 in \cite{Fer(02)}), an application of point $ii)$ of Lemma \ref{lm:merge} gives
\begin{equation} \label{eq:Merging3}
D_{dM}\big(\alpha_{n_1, \dots, n_P}^{(dM)}; \rho_{n_1, \dots, n_P}^{(dM)} \circ \mathcal{L}[-\mathrm{I}(\thetab_0), \dots, -\mathrm{I}(\thetab_0)] \big) \rightarrow 0
\end{equation}
as $\ n_1, \dots, n_P \rightarrow +\infty$, where $\alpha_{n_1, \dots, n_P}^{(dM)}$ denotes the probability distribution of the random vector 
$$
\Big(\sqrt{n_1 + \dots + n_G}\ \big(\thetab_{n_1, \dots, n_G}^{\ast} - \hat{\thetab}_{n_1, \dots, n_G}\big), \dots, \sqrt{n_M + \dots + n_P}\ \big(\thetab_{n_M, \dots, n_P}^{\ast} - \hat{\thetab}_{n_M, \dots, n_P}\big)
\Big) \ . 
$$
In addition, this argument entails also the tightness of the family of probability distributions $\{\alpha_{n_1, \dots, n_P}^{(dM)}\}_{n_1, \dots, n_P \geq 1}$. To conclude, one resorts to \eqref{eq:start1} and invokes
point $i)$ of Lemma \ref{lm:merge}, taking account that $\hat{\mathbb{Q}}_{n_i + \dots + n_{i + G - 1}}^{\ast\ast} \rightarrow \frac12 \mathrm{I}(\thetab_0)$ in probability as $n_i, \dots, n_{i + G - 1} \rightarrow +\infty$ (see again the proof of Theorem 22 in \cite{Fer(02)}), to get $D_M\big(\eta_{n_1, \dots, n_P}^{(M)}; \overline{\eta}_{n_1, \dots, n_P}^{(M)} \big) \rightarrow 0$
as $\ n_1, \dots, n_P \rightarrow +\infty$, where $\overline{\eta}_{n_1, \dots, n_P}^{(M)}$ stands for the probability law of the random vector having $i$-th component equal to
$(n_i + \dots + n_{i + G - 1})\ ^t\big(\thetab_{n_i, \dots, n_{i + G - 1}}^{\ast} - \hat{\thetab}_{n_i, \dots, n_{i + G - 1}}\big) \mathrm{I}(\thetab_0) \big(\thetab_{n_i, \dots, n_{i + G - 1}}^{\ast} - \hat{\thetab}_{n_i, \dots, n_{i + G - 1}}\big)$. Therefore, it remains only to combine Lemma \ref{lm:chi} with \eqref{eq:Merging1}-\eqref{eq:Merging2}-\eqref{eq:Merging3}.



\end{document}